\RequirePackage{fix-cm}
\documentclass[natbib,smallextended]{svjour3}       
\smartqed  
\usepackage{graphicx}
 \journalname{my journal}
\def\be{\begin{equation}}
\def\ee{\end{equation}}
\def\ba{\begin{eqnarray}}
\def\ea{\end{eqnarray}}
\def\go{\mathrel{\raise.3ex\hbox{$>$}\mkern-14mu\lower0.6ex\hbox{$\sim$}}}
\def\lo{\mathrel{\raise.3ex\hbox{$<$}\mkern-14mu\lower0.6ex\hbox{$\sim$}}}

\def\bB {{\bf B}}
\def\bk {{\bf k}}

\begin{document}

\title{Physics in Very Strong Magnetic Fields
}
\subtitle{Introduction and Overview}

\titlerunning{Physics in Strong Magnetic Fields}        

\author{Dong Lai}


\institute{D. Lai\at
              Department of Astronomy, Cornell University, Ithaca, NY 14853, USA \\
              Tel.: 1-607-255-4936\\
              Fax: 1-607-255-6918\\
              \email{dong@astro.cornell.edu}           
}

\date{Received: date / Accepted: date}

\maketitle

\begin{abstract}
This paper provides an introduction to a number of astrophysics
problems related to strong magnetic fields.  The first part deals with
issues related to atoms, condensed matter and high-energy processes in very
strong magnetic fields, and how these issues influence various aspects
of neutron star astrophysics. The second part deals with classical
astrophysical effects of magnetic fields: Even relatively ``weak''
fields can play a strong role in various astrophysical problems,
ranging from stars, accretion disks and outflows, to the formation and
merger of compact objects.  
\end{abstract}

\section{Introduction}
\label{intro}

The subject ``Physics in Very Strong Magnetic Fields'' is a very broad
one, and the title is also somewhat ambiguous. The first question one
may ask is: {\it How strong a magnetic field is ``Strong''?} The
answer to this question will depend on the objects one is dealing
with, on the issues one is interested in, and on whom one is talking
to.

In the following, we will first review issues of strong magnetic
fields from a general physics point of view and discuss how these
issues may relate to some aspects of neutron star astrophysics. This
focus on neutron stars reflects that fact that neutron stars are
endowed with the strongest magnetic fields in the universe where
fundamental strong-field physics can play an important role. It also
reflects the author's own research interest on the subject. For most
other astrophysics problems, covering a wide range of sub-fields of
astrophysics, magnetic fields are essentially classical, i.e., we are
essentially dealing with Maxwell equations. We will discuss why such
``weak'' magnetic fields can be considered strong, and how such fields
play an important role in various astrophysics contexts, ranging from
stars and star formation, to disks and outflows, and to stellar
megers.

\section{Atomic and Molecular Physics}
\label{sec:1}

When studying matter in magnetic fields, the natural (atomic) unit
for the field strength, $B_0$, is set by equating the electron cyclotron
energy $\hbar\omega_{ce}$ to the characteristic atomic energy
$e^2/a_0=2\times 13.6$~eV (where $a_0$ is the Bohr radius),
or equivalently by $\hat R=a_0$, where $\hat R=(\hbar c/eB)^{1/2}$
is the cyclotron radius of the electron. Thus it is convenient to
define a dimensionless magnetic field strength $b$ via
\be
b\equiv {B\over B_0};\qquad B_0={m_e^2e^3c\over\hbar^3}=2.3505\times 10^9\,
{\rm G}.
\label{eqb0}\ee
For $b\gg 1$, the cyclotron energy $\hbar\omega_{ce}$ is much larger
than the typical Coulomb energy, so that the properties of atoms,
molecules and condensed matter are qualitatively changed by the
magnetic field.  In such a strong field regime, the usual perturbative
treatment of the magnetic effects (e.g., Zeeman splitting of atomic
energy levels) does not apply. Instead, the Coulomb forces act as a
perturbation to the magnetic forces, and the electrons in an atom
settle into the ground Landau level. Because of the extreme
confinement ($\hat R\ll a_0$) of the electrons in the transverse
direction (perpendicular to the field), the Coulomb force becomes much
more effective in binding the electrons along the magnetic field
direction. The atom attains a cylindrical structure. Moreover, it is
possible for these elongated atoms to form molecular chains by
covalent bonding along the field direction. Interactions between the
linear chains can then lead to the formation of three-dimensional
condensates (see Lai 2001; Harding \& Lai 2006 for review).

{\bf (i) Atoms:} For $b\gg 1$, the H atom is elongated and squeezed,
with the transverse size (perpendicular to $\bB$) $\sim \hat R=a_0/b^{1/2}           
\ll a_0$ and the longitudinal size $\sim a_0/(\ln b)$. Thus the ground-state
binding energy $|E|\simeq 0.16\,(\ln b)^2$(au) (where 1~au =
27.2~eV; the factor
$0.16$ is an approximate number based on numerical calculations).
Thus $|E|=160,540$~eV at $B=10^{12}, 10^{14}$G respectively.
In the ground state, the guiding center of the electron's
gyro-motion coincides with the proton. The excited states of the
atom can be obtained by displacing the guiding center away from the proton;
this corresponds to $\hat R\rightarrow R_s=(2s+1)^{1/2}\hat R$ (where
$s=0,1,2,\cdots$). Thus $E_s\simeq -0.16\,\left\{\ln [b/(2s+1)]                      
\right\}^2$(au).

We can imagine constructing a multi-electron atom (with Z electrons)
by placing electrons at the lowest available energy levels of a
hydrogenic ion. The lowest levels to be filled are the tightly bound
states with $\nu=0$ (zero node in the wavefunction along the field direction). 
When $a_0/Z \gg \sqrt {2 Z-1} {\hat R}$, i.e.,
$b \gg 2 Z^3$, all electrons settle into the tightly bound levels with
$s=0,1,2,\cdots,Z-1$. Reliable values for the energy of a multi-electron
atom for $b\gg 1$ can be calculated using the Hartree-Fock method
or density functional theory, which takes into account the
electron-electron direct and exchange interactions
in a self-consistent manner.

{\bf (ii) Molecules and Chains:}
In a strong magnetic field, the mechanism of forming molecules is
quite different from the zero-field case. The spins of the electrons
in the atoms are aligned anti-parallel to the magnetic field, and thus
two atoms in their ground states do not bind together according to the
exclusion principle.  Instead, one H atom has to be excited to the
$s=1$ state before combining (by covalent bond) with another atom in
the $s=0$ state.  Since the ``activation energy'' for exciting an
electron in the H atom from $s$ to $(s+1)$ is small, the resulting
H$_2$ molecule is stable.  Moreover, in strong magnetic fields, stable
H$_3$, H$_4$ etc. can be formed in the similar manner. The
dissociation energy of the molecule is much greater than the $B=0$
value: e.g., for H$_2$, it is 40,350~eV at $10^{12},10^{14}$~G
respectively. A highly magnetized molecule exhibits excitation
levels much different from a $B=0$ molecule.

{\bf (iii) Neutron Star Atmospheres and Radiation:}
An important area of research where the atomic physics in strong
magnetic fields plays an important role is the study of neutron star
(NS) atmospheres and their radiation
(see Potekhin et al.~2014 for more details).
Thermal, surface emission from isolated
NSs can potentially provide invaluable information on the physical
properties and evolution of NS (equation of state at super-nuclear
densities, superfluidity, cooling history, magnetic field, surface
composition, different NS populations).
In recent years, considerable
observational resources (e.g. {\it Chandra} and {\it XMM-Newton}) have
been devoted to such study.
For example, the spectra of a number of
radio pulsars (e.g., PSR~B1055-52, B0656+14, Geminga and Vela) have
been observed to possess thermal components that can be attributed to
emission from NS surfaces and/or heated polar caps.
Phase-resolved spectroscopic observations are becoming
possible, revealing the surface magnetic field geometry and emission
radius of the pulsar.
A number of compact sources in
supernova remnants have been observed, with spectra consistent with thermal emission from
NSs, and useful constraints on NS cooling
physics have been obtained.
Surface X-ray emission has also been detected from a number of
SGRs and AXPs. Fits to the quiescent magnetar spectra with blackbody or with
crude atmosphere models indicate that the thermal X-rays can be
attributed to magnetar surface emission at temperatures of
(3--7) $\times 10^6$~K .
One of the intriguing puzzles is
the absence of spectral features (such as ion cyclotron line
around 1~keV for typical magnetar field strengths) in the observed
thermal spectra. Clearly, detailed observational and theoretical studies of
surface emission can potentially reveal much about
the physical conditions and the nature of magnetars.

Of particular interest are the seven
isolated, radio-quiet NSs (so-called ``dim isolated NSs''; see
van Kerkwijk \& Kaplan 2007; Haberl 2007).
These NSs share the common property that their spectra
appear to be entirely thermal, indicating
that the emission arises directly from the NS atmospheres,
uncontaminated by magnetospheric processes.
Thus they offer the best hope for inferring the precise values of the
temperature, surface gravity, gravitational redshift and magnetic field
strength. The true nature of these sources, however, is unclear at present:
they could be young cooling NSs, or NSs kept hot by accretion
from the ISM, or magnetars and their descendants.
Given their interest, these isolated NSs have been intensively studied by
deep {\it Chandra} and {\it XMM-Newton} observations.
While the brightest of these, RX J1856.5-3754, has a featureless
spectrum remarkably well described by a
blackbody, absorption lines/features
at $E\simeq 0.2$--$2$~keV have been detected in six other sources,
The identifications of these features, however, remain uncertain,
with suggestions ranging from cyclotron lines to atomic transitions of H,
He or mid-Z atoms in a strong magnetic field (see Ho \& Lai 2004; 
Ho et al.~2008;
Potekhin et al.~2014).
Another puzzle concerns the optical emission: For four sources,
optical counterparts have been identified, but the optical flux is larger
(by a factor of $4$-$10$) than the extrapolation from the black-body fit to
the X-ray spectrum. Clearly, a proper understanding/interpretation of these
objects requires detailed NS atmosphere modeling which includes careful treatments
of atomic physics in strong magnetic fields.

\section{Condensed Matter Physics}
\label{sec:3}

Several aspects of condensed matter physics in strong magnetic fields
play an important role in neutron star astrophysics.

{\bf (i) Cohesive Property of Condensed Matter:} Continuing
our discussion of atoms/molecules in strong magnetic fields,
as we add more atoms to a H molecular chain,
the energy per atom in a H$_n$ molecule saturates, becoming independent
of $n$. We then have a 1D metal. Chain-chain interactions then lead to 3D
condensed matter. The binding energy of magnetized condensed matter at zero pressure 
can be estimated using the uniform electron gas model.
Balancing the electron kinetic (zero-point) energy and the Coulomb
energy in a Wigner-Seitz cell (containing one nucleus and $Z$ electrons),
we find that the energy per unit cell is of order $E\sim -Z^{9/5}b^{2/5}$.
The radius of the cell is $R\sim Z^{1/5}b^{-2/5}$, corresponding to
the zero-pressure density $\simeq 10^3AZ^{3/5}B_{12}^{6/5}$~g~cm$^{-3}$
(where $A$ is the mass number of the ion).

Although the simple uniform electron gas model and its Thomas-Fermi
type extensions give a reasonable estimate for the binding energy for
the condensed state, they are not adequate for determining the
cohesive property of the condensed matter.  In principle, a
three-dimensional electronic band structure calculation is needed to
solve this problem. 
The binding energies of 1D chain for some elements have been obtained
using Hartree-Fock method (Neuhauser \etal~1987; Lai
\etal~1992). Density functional theory has also been used to calculate
the structure of linear chains in strong magnetic fields (Jones 1986;
Medin \& Lai 2006a,b).  Numerical calculations carried out so far
indicate that for $B_{12}=1-10$, linear chains are unbound for large
atomic numbers $Z\go 6$. In particular, the Fe chain is unbound
relative to the Fe atom; therefore, the chain-chain interaction must
play a crucial role in determining whether the 3D zero-pressure Fe
condensed matter is bound or not.  However, for a sufficiently large
$B$, when $a_0/Z\gg \sqrt{2Z+1}{\hat R}$, or $B_{12}\gg 100(Z/26)^3$,
we expect the Fe chain to be bound in a manner similar to the H chain
or He chain (Medin \& Lai 2006a,b).  The cohesive property of
magnetized condensed matter is important for understanding the
physical condition of the “polar gap” and particle acceleration 
in pulsars (Medin \& Lai 2007).

{\bf (ii) Phase Diagram and Equation of State:} Given the energies of
different bound states of a certain element, one can determine the
phase diagram as a function of the field strength $B$ and
temperature. This is relevant to the outmost layer of neutron stars
(NSs). For a given $B$, there is a critical temperature below which
the phase separation will occur, and the NS surface may be in a
condensed state, with negligible gas above it. Some isolated NSs with
low surface temperatures may be in such a state (see van Adelsberg et
al.~2005; Medin \& Lai 2007).

Beyond zero-pressure density, the Coulomb interaction can be neglected, 
and the effects of magnetic field on the equation of state of
matter depend on $B$, $\rho$ and $T$. We can define a
critical ``magnetic density'', below which only the ground Landau
level is populated (at $T=0$), given by 
\be
\rho_B=0.802\,Y_e^{-1}
b^{3/2}~{\rm g~cm}^{-3}=7.04\times 10^3\,Y_e^{-1} B_{12}^{3/2}
~{\rm g~cm}^{-3},
\label{eqrhob}\ee
where $Y_e=Z/A$ is the number of electrons per baryon. We can also
define critical ``magnetic temperature'', 
\be 
T_B\simeq
{\hbar\omega_{ce}\over k_B}\left({m_e\over m_e^\ast}\right)
=1.34\times 10^8\,B_{12}\,(1+x_F^2)^{-1/2}~{\rm K}, 
\ee 
where $m_e^\ast=\sqrt{m_e^2+(p_F/c)^2}=m_e\sqrt{1+x_F^2}$.  There are three
regimes characterizing the effects of Landau quantization on the
thermodynamic properties of the electron gas: 

(a) $\rho\lo\rho_B$ and
$T\lo T_B$: In this regime, the electrons populate mostly the ground
Landau level, and the magnetic field modifies essentially all the
properties of the gas. The field is sometimes termed ``strongly
quantizing''.  For example, for degenerate, nonrelativistic electrons
($\rho<\rho_B$ and $T\ll T_F\ll m_ec^2/k_B$, where $T_F$ is the Fermi temperature), 
the pressure is $P_e=(2/3)n_eE_F\propto B^{-2}\rho^3$.  This should be compared with
the $B=0$ expression $P_e\propto\rho^{5/3}$.  Note that for
nondegenerate electrons ($T\gg T_F$), the classical ideal gas equation
of state, $P_e=n_ek_BT$, still holds in this ``strongly quantizing'' regime.

(b) $\rho\go\rho_B$ and $T\lo T_B$: In this regime, the electrons are
degenerate, and populate many Landau levels but the level
spacing exceeds $k_BT$. The magnetic field is termed ``weakly                   
quantizing''.  The bulk properties of the gas (e.g., pressure and
chemical potential) are only slightly affected by such magnetic
fields. However, the quantities determined by thermal electrons near
the Fermi surface show large oscillatory features as a function of
density or magnetic field strength. These de Haas - van Alphen type
oscillations arise as successive Landau levels are occupied with
increasing density (or decreasing magnetic field). 
With inceasing $T$, the oscillations
become weaker because of the thermal broadening of the Landau levels;
when $T\go T_B$, the oscillations are entirely smeared out, and the
field-free results are recovered.

(c) $T\go T_B$ or $\rho\gg\rho_B$: In this regime, many
Landau levels are populated and the thermal widths of the Landau
levels ($\sim k_BT$) are higher than the level spacing. The magnetic
field is termed ``non-quantizing'' and does not affect the
thermodynamic properties of the gas.

{\bf (iii) Transport Properties:} 
A strong magnetic field can significantly affect the transport
properties and thermal structure of a neutron star crust.
Even in the regime where the magnetic quantization effects are small
($\rho\gg\rho_B$), the magnetic field can still greatly modify
the transport coefficients (e.g., electric conductivity and heat
conductivity). This occurs when the effective
gyro-freqyency of the electron, $\omega_{ce}^\ast=eB/(m_e^\ast c)$,
where $m_e^\ast=\sqrt{m_e^2+(p_F/c)^2}$, is much larger than the
electron collision frequency $1/\tau_0$.
When $\omega_{ce}^\ast\tau_0\gg 1$, the electron
heat conductivity perpendicular to the magnetic field,
$\kappa_\perp$, is suppressed by
a factor $(\omega_{ce}^\ast\tau_0)^{-2}$. In this classical
regime, the heat conductivity along the field, $\kappa_\parallel$,
is the same as the $B=0$ value. In a quantizing magnetic field,
the conductivity exhibits oscillatory behavior of the
de Haas - van Alphen type. On average, the longitudinal
conductivity is enhanced relative to the $B=0$ value due to quantization.
The most detailed calculations of the electron transport
coefficients of magnetized neutron star envelopes
are due to Potekhin (1999), where earlier references
can be found (see Potekhin et al.~2014 for more details).

The thermal structure of a magnetized neutron star envelope
has been studied by many authors (see Potekhin et al.~2014 for review).
In general, a normal magnetic field reduces the thermal insulation
as a result of the (on average) increased $\kappa_\parallel$ due to
Landau quantization of electron motion, while
a tangential magnetic field (parallel to the stellar surface)
increases the thermal insulation of the envelope because the
Lamor rotation of the electron significantly reduces
the transverse thermal conductivity $\kappa_\perp$.
A consequence of the anisotropic heat transport is that for a given
internal temperature of the neutron star, the surface temperature
is nonuniform, with the magnetic poles hotter and the magnetic
equator cooler (see, e.g., Shabaltas \& Lai 2012 for a recent application).

\section{High Energy Physics: QED in Strong Mganetic Fields}
\label{sec:5}

In superstrong magnetic fields, a number of quantum-electrodynamic (QED)
processes are important. A well-known one is single-photon pair production,
$\gamma \rightarrow e^+ + e^-$. This process is forbidden at zero-field, but 
is allowed for $B\neq 0$, and is one of the dominant channels for pair cascade
in pulsar magnetospheres (Srurrock 1971; Medin \& Lai 2010). Another process is 
photon splitting, $\gamma\rightarrow\gamma+\gamma$, which can attain appreciable
probability for sufficiently strong fields. The critical QED field strength is 
set by $\hbar\omega_{ce}=m_ec^2$, i.e.,
\be
B_Q={m_e^2c^3\over e\hbar}=4.414\times 10^{13}~{\rm G}.
\ee
Above $B_Q$, many of these QED effects become important.

A somewhat surprising strong-field QED effect is vacuum
polarization, which makes even an ``empty'' space birefringent for
photons propagating through it. This can significantly affect radiative transfer 
in neutron star atmospheres and the observed spectral and x-ray polarization signals
even for modest field strengths. We discuss this issue below.

The magnetized plasma of a NS atmosphere is birefringent.
A X-ray photon, with energy $E\ll E_{ce}=\hbar\omega_{ce}=1.16B_{14}$~MeV
[where $B_{14}=B/(10^{14}\,{\rm G})$],
propagating in such a plasma can be
in one of the two polarization modes: The ordinary mode (O-mode) has its
electric field ${\bf E}$ oriented along the $\bB$-$\bk$ plane
($\bk$ is along direction of propagation),
while the extraordinary mode (X-mode) has its ${\bf E}$ perpendicular to
the $\bB$-$\bk$ plane. Since charge particles cannot move freely across
the magnetic field, the X-mode photon opacity
(e.g., due to free-free absorption or electron scattering)
is suppressed compared to the zero-field value,
$\kappa_X\sim (E/E_{Be})^2\kappa_{(B=0)}$, while the O-mode
opacity is largely unchanged, $\kappa_O\sim \kappa_{(B=0)}$ (e.g., Meszaros 1992).
Vacuum polarization can change this picture in an essential way.
In the presence of a strong magnetic field,
vacuum itself becomes birefringent due to virtual $e^+e^-$ pairs.
Thus in a magnetized NS atmosphere, both the plasma and vacuum polarization
contribute to the dielectric tensor of the medium.
The vacuum polarization
contribution is of order $10^{-4}(B/B_Q)^2f(B)$ (where
$B_Q=m_e^2c^3/e\hbar=4.414\times 10^{13}$~G, and $f\sim 1$ is a slowly
varying function of $B$), and is quite small unless $B\gg B_Q$.
However, even for ``modest'' field strengths, vacuum polarization can have
a dramatic effect through a ``vacuum resonance'' phenomenon.
This resonance arises when the effects of
vacuum polarization and plasma on the polarization of the photon modes
``compensate'' each other. For a photon of energy $E$ (in keV), the vacuum
resonance occurs at the density
$\rho_V\simeq 0.964\,Y_e^{-1}B_{14}^2E^2 f^{-2}~{\rm g~cm}^{-3}$,
where $Y_e$ is the electron fraction (Lai \& Ho 2002).
Note that $\rho_V$ lies in the range of the typical densities
of a NS atmosphere.
For $\rho\go\rho_V$ (where the plasma effect dominates the dielectric
tensor) and $\rho\lo\rho_V$ (where vacuum polarization dominates), the
photon modes are almost linearly polarized --- they are the usual
O-mode and X-mode described above; at $\rho=\rho_V$, however, both
modes become circularly polarized
as a result of the ``cancellation'' of the plasma and vacuum polarization
effects 
When a photon propagates outward
in the NS atmosphere, its polarization state will evolve adiabatically
if the plasma density variation is sufficiently gentle. Thus the photon
can convert from one mode into another as it traverses the
vacuum resonance. The conversion probability $P_{\rm conv}$ depends
mainly on $E$ and atmosphere density gradient; for a typical atmosphere
density scale height ($\sim 1$~cm), adiabatic mode conversion requires
$E\go 1$-2~keV (Lai \& Ho 2003a).
Because the O-mode and X-mode have vastly different opacities,
the vacuum polarization-induced mode conversion can significantly affect
radiative transfer in magnetar atmospheres.
In particular,
the effect tends to deplete the high-energy tail of the thermal spectrum
(making it closer to blackbody)
and reduce the width of the ion cyclotron line or other spectral lines
(Ho \& Lai 2003,2004; Lai \& Ho 2003a; van Adelsberg \& Lai 2006).
It is tempting to suggest that the absence of lines in the observed thermal
spectra of several AXPs is a consequence of the vacuum polarization
effect at work in these systems.  

We also note that even for ``ordinary'' NSs (with $B\sim
10^{12}$-$10^{13}$~G), vacuum resonance has a profound effect on the
polarization signals of the surface emission; this may provide a
direct probe of strong-field QED in the regime inaccessible at
terrestrial laboratories (Lai \& Ho 2003b; Wang \& Lai 2009; see Lai 2010 for 
a review). Such
polarization signals will be of interest for future X-ray polarimetry
detectors/missions.

Finally, magnetic fields can modify neutrino processes that take place in
neutron stars. For example, in proto-neutron stars with sufficiently strong
B-fields, the neutrino cross sections and emission rates, as well as their
angular dependences, can be affected, and these can contribute to
the natal velocity kick imparted to the neutron star (e.g.,
Arras \& Lai 1999a,b; Maruyama et al.~2014).

\section{``Classical'' Astrophysics}
\label{sec:6}

For most areas of astrophysics, magnetic fields are
``classical''. That is, we are dealing with Maxwell's equations, MHD and
classical plasma physics.  The quantization, microscopic effects
discussed previous sections are not relevant. Nevertheless, these
classical magnetic field effects are important, interesting and rich.
We will highlight some of these in the following.

\subsection{Clouds, Stars and Compact Objects}

The first effect of ``classical'' magnetic fields is that they can influence
the equlibrium of bound objects via the so-called magnetic Virial theorem.
For a spherical cloud or star of mass $M$ and mean radius $R$, static equilibrium 
requires that  the ratio of the magnetic energy and gravitational energy be 
less than unity, i.e.,
\be
{E_{\rm mag}\over E_{\rm grav}}\sim {B_{\rm in}^2R^3/6\over GM^2/R}
\sim {1\over 6\pi^2G}\left({\Phi\over M}\right)^2\lo 1,
\label{eq:emag}\ee
where the second equality assumes that the dominant internal magnetic field
takes form of a large-scale poloidal field, and $\Phi=\pi R^2 B_{\rm in}$ is the 
magnetic flux threading the cloud.

In the context of star formation, clouds (cores) with $E_{\rm mag}/E_{\rm grav}
\go 1$ cannot collapse on a dynamical timescale, but require ambipolar diffusion
to eliminate the magnetic flux. This process is perhaps relevant for
the formation of low-mass stars (e.g., Shu et al.~1999), although
in recent years the roles of turbulence in the molecular clouds
have been recognized (McKee \& Ostriker 2007).

For neutron stars (with $M\simeq 1.4M_\odot$ and $R\simeq 10$~km), 
equation (\ref{eq:emag}) implies $B_{\rm in}\lo 10^{18}$~Gauss.
This is the maximum field strength achievable in all astrophysical objects.

What do we know observationally about magnetic fields of isolated
neutron stars?  
For radio pulsars, the dipole magnetic fields are
inferred indirectly from the measured $P$ and $\dot P$ (rotation
period and period derivative), and the assumption that the spindown is
due to magnetic dipole radiation. For most ``regular'' pulsars, the magnetic fields
thus derived lie in the range of $10^{12-13}$~G. A smaller population, so-called
``millisecond pulsars'', have fields in the range of $10^{8-9}$~G. 
How such a ``weak'' field evolves from the regular field of $10^{12-13}$~G
remains unclear (see Payne \& Melatos 2004). In recent years, a number of
``High-B Radio Pulsars'' have also been found: these have $B\sim 10^{14}$~G,
comparable to magnetars.

Magnetars are neutron stars powered by energy dissipation of magnetic
fields.  They usually have dipole fields (as inferred from $P$ and
$\dot P$ based on x-ray timing) of $B\go 10^{14}$~G. Interestingly, a
number of low-field ($\sim 10^{13}$~G) magnetars have also been found
recently (Rea et al.~2010), although the internal fields could be
higher. Indeed, there is growing evidence that there exist hidden
magnetic fields inside neutron stars. This is the case for the neutron
star in Kes 79 SNR: It has a dipole field of $3\times 10^{10}$~G, but
the internal field buried inside its crust could be larger than
$10^{14}$~G, based on its observed large x-ray pulse fraction of $60\%$
(Halpern \& Gotthelf 2010; Shabaltas \& Lai 2012; Vigano et al.~2013).
In the case of SGR 0418+5729, the dipole field is less than a few times
$10^{12}$~G, but internal field could be much stronger (Turolla et al.~2011).

Another way to assess whether a magnetic field is ``strong'' is to
look at the energetics. For magnetars, even in quiescence, the x-ray
luminosity is $L\sim 10^{34-36}$~erg~s$^{-1}$, much larger than the
spindown luminosity ($I\Omega\dot\Omega$). The giant flares of the
three SGRs indicate that a much larger internal field is possible. For
example, the December 2004 flare of SGR 1806-20 has a total energy of
$10^{46}$~erg, suggesting an internal field of at least a few times
$10^{14}$~G.

What is the origin of such strong magentic fields? It is intriguing to
note that (Reisenegger 2013) for upper main-sequence stars (radius
$10^{6.5}$~km), white dwarfs ($10^4$~km) and neutron stars ($10$~km),
the maximum observed magnetic fields ($10^{4.5}$~G, $10^9$~G and
$10^{15}$~G respectively) all correspond to similar maximum magnetic
flux $\Phi_{\rm max}=\pi R^2 B_{\rm max}\sim
10^{17.5-18}$~G~km$^2$. This seems to suggest a fossil origin of the
strongest magnetic fields. However, recent observations of magnetic
white dwarfs (and their populations in binaries) indicate the strong
magnetic fields ($\go$ a few MG) of white dwarfs originate from binary
mergers (Wickramasinghe et al.~2014). So perhaps the strongest magentic 
fields found in magnetars is the result of dynamo action in the proto-neutron
star phase (Thompson \& Duncan 1993). In any case, since $E_{\rm mag}/E_{\rm grav}\lo
10^{-6}$ (assuming no significant hidden magnetic fields), these magnetic fields
have a negligible effect on the global static equilibrium of the star.

\subsection{Stellar Envelopes and ``Outside''}

Although astrophysically observed magnetic fields have a negligible 
effect on the global equilibrium of a star, they can strongly influence
the local ``static'' equilibrium of stellar envelopes. A notable example
is neutron star (NS) crust. Because of the evolution of crustal magnetic fields
due to a combination of Hall drift and Ohmic diffusion,
the NS crust can break (e.g. Pons \& Perna 2011).
This occurs when $B^2/(8\pi)\go \mu \theta_{\rm max}$ (where $\mu$ is the shear
modulus and $\theta_{\rm max}$ is the maximum strain of the crust), or
$B\go 2\times 10^{14}(\theta_{\rm max}/10^{-3})^{1/2}$~G. The consequences of the 
crustal breaking (and its manifestations such as magnetar flares) 
are not clear. They depend on whether 
the breaking is fast or slow. The energy release and whether the energy can get 
out of the NS are also uncertain (see Link 2014; Beloborodov \& Levin 2014).

Of course, outside the star, even a ``weak'' magnetic field can be
quite ``strong'' and dominates the dynamics of the flow. Such
magnetically dominated region is relevant to the
magnetic braking of stars. In the case of radio pulsars, the
electrodynamics and physical processes in the magnetosphere are
ultimately responsible for most of the observed phenomena of pulsars.
In recent years, there has been significant progress in ab initio calculations of
pulsar magnetospheres (e.g. Tchekhovskoy et al.~2013), although it remains
unclear whether the current theoretical approach can adequately explain some
of the enigmatic pulsar phenomena (such as mode-switching in radiation; 
e.g. Hermsen et al.~2013). The magnetospheres of magnetars have also been studied:
Unlike radio pulsars, the closed field line regions play an important role
(e.g. Thompson et al.~2002; Beloborodov 2013).

Finally, further away from pulsars, we have pulsar wind nebulae, where pulsar
wind impinges upon a supernova remnant, creating a broad spectrum of
non-thermal radiation (from radio to gamma rays). The ultimate source of this
radation is the pulsar's rotational energy, and magnetic field plays an important
role in making such a ``transfer of energy'' possible (e.g. Amato 2013).

\subsection{Accretion Disks}

Magnetic fields play a number of important roles in accretion disks.
First, we have magnetically dominated disks. These occur when
$B^2/(8\pi)\go \rho v_k^2/2$, where $v_k$ is the Keplerian velocity of
the disk and $\rho$ is the density. In the last few years, a number of
studies have shown that the innermost region of a disk around a black hole
may accumulate large magnetic flux, and relativistic jets can be
generated through the Blandford-Znajek process (e.g. McKinney et
al.~2012). However, a physical understanding of the state transition
and jets (both steady and episodic) from black-hole x-ray binaries
remains elusive (Fender \& Belloni 2012; Yuan \& Narayan 2014).

Outflows can be launched from disks with large-scale
super-thermal magnetic
fields (at the disk surface), $B^2/(8\pi)\go \rho c_s^2/2$ (where
$c_s$ is the sound speed). Such magnetocentrifugal winds/outflows (a la 
Blandford-Payne) may occur in x-ray binaries (in the thermal
state) and in protostars (see Konigl \& Pudritz 2000 for a review).

Such large-scale strong magnetic fields are unlikely to be produced in
the disk by dynamo processes, and must be advected inward from large
radii. This is an important issue that has received a lot of theoretical
attention. The radial inward advection speed is $|u_r|\sim \nu/r$, and the outward 
Ohmic diffusion speed is $u_{\rm diff}\sim (\eta/H)(B_r/B_z)$, where 
$\nu$ is the disk viscosity, $\eta$ is the magentic diffusivity and $H$ is the
disk thickness. Clearly, the net outcome depends on the magnetic Prandtl number
$P_r=\nu/\eta$, which is typically of order unity (based on local MRI
turbulence simulations; Lesur \& Longaretti 2009). Recent work has emphasized the 
importance of proper treatment of vertical structure of the disk (e.g., the
electric conductivity is higher at the disk surface, so the field
advection is faster than mass advection; Lovelace et al.~2009;
Guilet \& Ogilvie 2013).

Finally, even ``weak'' sub-thermal magnetic fields can play an
important role in accretion disks. It is now well-established that for
most astrophysical disks, MRI (magneto-rotational instability) driven
turbulence is responsible for generating the anomalous viscosity
needed for accretion to proceed (Balbus \& Hawley 1998). It is also
recognized that the strength of the turbulence depends on the net
vertical field threading the disk (Hawley et al.~1995; Simon et
al.~2013). Recent works have emphaszied the roles of non-ideal MHD
effects in surpressing turbulence in proto-planetary disks (Bai \& Stone 2013;
Bai 2014).

\subsection{Disk Accretion onto Magnetic Stars}

Disk accretion onto magnetic central objects occurs in a variety of
astrophysical contexts, ranging from classical T Tauri stars,
and cataclysmic variables (intermediate polars),
to accretion-powered X-ray pulsars. The basic picture of disk-magnetosphere
interaction is well known: The stellar magnetic field disrupts the
accretion flow at the magnetospheric boundary and funnels the plasma
onto the polar caps of the star or ejects it to infinity.
The magnetosphere boundary is located where
the magnetic and plasma stresses balance,
\be
r_m =\xi \left({\mu^4\over GM\dot M^2}\right)^{1/7},
\ee
where $M$ and $\mu$ are the mass and magnetic moment of the central
object, $\dot M$ is the mass accretion rate and $\xi$ is a
dimensionless constant of order 0.5-1. 
Roughly speaking, the funnel flow occurs when $r_m$ is less than the corotation
radius $r_c$ (where the disk rotates at the same rate as the
star). For $r_m\go r_c$, centrifugal forces may lead to ejection of
the accreting matter (``propeller'' effect).

Over the years, numerous theoretical studies have been devoted to
understanding the interaction between accretion disks and magnetized
stars. Many different models have been developed (see
Lai 2014 for a review).
In parallel to these theoretical studies, there have been
many numerical simulations, with increasing sophistication.
These simulations are playing an important role in
elucidating the physics of magnetosphere-disk interaction in various
astrophysical situations (see Romanova et al.~2014 and Zanni 2014 for review).

The problem of magnetosphere-disk interaction has many applications:
(i) Rotation rate of protostars: Many protostars are found to have
rotation rates about $10\%$ of breakup. Magnetosphere spin equilibrium
($r_m$ equals the corotation radius) has long been suggested, although
magnetosphere/stellar winds may also play a role (Gallet \& Bouvier
2013).  (ii) Spinup/spindown of accreting x-ray pulsars: Many x-ray
pulsars have been observed to exhibit changing spinup and spindown
behavior over timescales of years. For example, 4U1626-67 is an
accreting pulsar with spin period 7.66~s. The clean spinup before
1990.6 was followed by a clean spindown, and another spinup phase
starting 2008.2. The spindown/spinup transition lasted 150 days. 
Understanding this spindown/spinup behavior and its correlation with
the accretion rate remains an outstanding unsolved problem.

When the stellar field lines penetrate some region of the disk, they
provide a linkage between the star and the disk. These field lines are
twisted by differential rotation between the stellar rotation
$\Omega_s$ and the disk rotation $\Omega(r)$, generating a toroidal
field.  However, when the toroidal field becomes comparable to the
poloidal field, the flux tube connecting the star and the disk will
start expanding. This field inflation is driven by the pressure
associated with the toroidal field. As the fields open up, the
star-disk linkage is broken. Such field-opening behavior has been
well-established through theoretical studies and numerical simulations
in the contexts of solar flares and accretion disks (Lovelace et
al.~1995).  Given this constraint on the toroidal twist, steady-state
disk-star linkage is possible only very near corotation.  In general,
we should expect a quasi-cyclic behavior, involving several stages:
(1) The stellar field penetrates the inner region of the disk; (2) The
linked field lines are twisted; (3) The resulting toroidal fields
drive field inflation; (4) Reconnection of the inflated field restores
the linkage.  The whole cycle then repeats (see Aly \& Juijpers 1990;
Uzdensky et al.~2002).  This quasi-cyclic behavior may be relevant to
QPOs observed in low-mass X-ray binaries (see van der Klis 2006 for a
review; Shirakawa \& Lai 2002a,b) and other systems, as well as give
rise to episodic outflows and winds (Zanni \& Ferreira 2013).

Finally, we note that in the standard picture of magnetic star-disk
interaction, it is usually assumed that the stellar spin axis is
aligned with the disk axis (the disk normal vector). This seems
reasonable since the star may have gained substantial angular momentum
from the accreting gas in the disk. However, magnetic interaction
between the star and the inner region of the disk may (if not always)
change this simple picture (Lai 1999,2003), giving rise to 
stellar spin - disk misalignment. This has application to spin-orbit
misalignment in exoplanetary systems (Lai et al.~2011; Foucart \& Lai 2011).

\subsection{Magnetic Fields in the Formation of Compact Objects}

In the ``standard'' picture of core-collapse supernovae leading to the
formation of neutron stars, neutrino heating behind the stalled
accretion shock, plus various hydrodynamical instabilities, are
responsible for the explosion. Magnetic fields play a negligible role
in this picture. However, there is a long list of theoretical works
exploring the role of magnetic fields in supernovae (LeBlanc \& Wilson
1970; Bisnovatyi-Kogan et al.~1976; Moiseenko et al.~2006; Burrows et
al.~2007). The key requirement for the magnetic field
to play a role is that the pre-SN core must have sufficiently rapid
rotation -- this is rather uncertain observationally. 
A technical challenge is that if one starts out with a modest magnetic field,
and use MRI dynamo to amplify the field, it is important that the MRI scale
is resolved in the numerical code -- this is currently not achieved
unless the initial field is greater than $10^{15}$~G.

In general, newly formed magnetars can play two roles in supernovae.
(i) They can power the explosion if the initial spin period of the proto-neutron
star is less than $\sim 3$~ms and the magnetic field is $10^{15}$~G or higher
(Bodenheimer \& Ostriker 1974; Thompson et al.~2004).
(ii) For modest rotation period ($\sim 10$~ms), the released rotational energy 
does not affect the explosion itself, but can still impact the SN lightcurves
(since the spindown timescale, about days to weeks, is comparable to the photon
diffusion time through the remnant). Such energy injection may help explain some 
of the super-luminous SNe with $L\go 10^{44}$~erg~s$^{-1}$ (Kasen \& Bildsten 2010;
Woosley 2010). In this regard, it is of interest to note that many central compact 
objects in SNRs have been found to possess rather weak dipole fields 
($B\lo 10^{12}$~G) and slow rotation (period $\sim 0.1$~s),
although the internal fields may be much stronger.

Magnetic fields play an important role in the central engine of long
Gamma-Ray Bursts (GRBs). Two scenarios are often discussed: (1) With
rapid rotation, core collapse leads to the formation of a
hyper-accreting black hole. Neutrino heating and magnetic fields (via
Blandford-Znajek process) then lead to the production of relativistic
jets (Zhang et al.~2003).  (2) Core collapse leads to the
formation of millisecond magnetars, which power the GRB
outflows/jets. Recent observations of long-lasting ($\sim 10^4$~s)
x-ray emission/flares suggest that long-lasting central engine may be
needed for some GRBs (Kumar \& Zhang 2014). Also, the observed high
polarization in reverse-shock emission indicates that large-scale
magentic fields are present in the GRB jets (Mundell et al.~2013).

\subsection{Magnetic Fields in Merging Compact Binaries}

There are two types of merging compact binaries that are of great interest:
(1) NS/NS and NS/BH binaries: These produce gravitational waves that are 
detectable by LIGO/VIRGO and generate electromagnetic counterparts in the form
of short GRBs and kilo-novae.
(2) Compact WD/WD binaries: These produce various exotic outcomes (R CrB stars,
AM CVn binaries, and possibly accretion-induced collapse and SN Ia), and
generate low-frequency gravitational waves detectable by LISA/NGO.

In recent years, there have been significant progress in simulating (in full
General Relativity) the merger of NS/NS binaries (e.g. Shibata et al.~2006)
and NS/BH binaries (Foucart et al.~2013). Simulations with magnetic fields
are also becoming possible (Giacomazzo et al.~2011; Palenzuela et al.~2013),
although much remains to be understood.

One issue of great interest is the merger of the NS magnetospheres
prior to the merger of the stars. The combined binary system can behave
as a single unipolar inductor producing radio waves that may be
detectable (Hansen \& Lyutikov 2001), although this is highly
uncertain and detailed calculations are difficult. Nevertheless, a
robust upper limit of the energy dissipation power in the
magnetosphere that can be generated prior to NS merger can be obtained
(Lai 2012).  This upper limit indicates that the magnetospheric
dissipation will not affect the orbital decay rate (and the
gravitational waveform), although the prospect for radio detection remains
uncertain.

Another issue of interest is the production of magnetic fields during
NS/NS binary merger. Since the binary NSs cannot be spin-synchronized
(because of the rapid orbital decay in the last few minutes of the
binary lifetime), strong velocity shear is present when the two stars
touch each other: Kelvin-Helmholtz instability develops at the
interface, which may then lead to the generation of strong magnetic fields
(Price \& Rosswog 2006).  Recent studies, however, suggest the
dynamical impact of such magnetic fields may be limited to the shear
layer (Obergaulinger, Aloy \& Muller 2010), although the situation is
not entirely clear (see Giacomazzo et al.~2014). Finally, the magnetic field
in the merger remnant is of great importance. This situation is similar to
the remnant in core-collapse supernova: Can the initial (weak/modest) magnetic
field be amplified by differential rotation and MRI dynamo (Is MRI resolved
in the simulation)? How are winds/outflows/jets produced? Is a black-hole 
or millisecond magnetar formed in the merger remnant?

\begin{acknowledgements}
I thank ISSI for hosting the workshop on Strongest Magnetic Fields in the
Universe. I thank my collaborators and students who have helped me
understand various topics discussed in this review. I also thank K. Anderson 
and J. Zanazzi for reading parts of the first draft.
This work has been supported in part by NSF grants AST-1008245, 1211061,
and NASA grant NNX12AF85G.
\end{acknowledgements}



\end{document}